\documentclass[a4paper]{article}

\usepackage{amsfonts}
\usepackage{amsopn}
\usepackage{amsmath}
\usepackage{amsthm}
\usepackage{latexsym}
\usepackage{amssymb}
\usepackage{graphicx}
\usepackage{dsfont}

\newcommand{\PP}{{\mathbb{P}}}

\newcommand{\RR}{{\mathbb{R}}}

\newcommand{\nF}{{\mathcal{F}}}

\newcommand{\nM}{{\mathcal{M}}}
\newcommand{\nN}{{\mathcal{N}}}

\newcommand{\re}{{\mathrm{e}}}

\newcommand{\boldeta}{{\boldsymbol{\eta}}}

\newcommand{\balpha}{{\boldsymbol{\alpha}}}
\newcommand{\bbeta}{{\boldsymbol{\beta}}}
\newcommand{\bx}{{\boldsymbol{x}}}
\newcommand{\bP}{{\boldsymbol{P}}}
\newcommand{\by}{{\boldsymbol{y}}}

\newcommand{\btheta}{{\boldsymbol{\theta}}}

\renewcommand\d{\, \mathrm{d}}

\newcommand{\p}{\partial}
\newcommand\1{\mathds{1}}
\newcommand{\dbar}{\, \mathrm{d}\llap{\raisebox{0.85ex}{$\scriptstyle-\!$}}}
\renewcommand{\phi}{\varphi}

\newcommand{\norm}[1]{\left\|#1\right\|}
\newcommand{\dgl}[1]{\left\{\begin{aligned}
	#1
	\end{aligned}\right.}
\newcommand{\abs}[1]{\left|#1\right|}

\newcommand{\menge}[1]{\left\lbrace #1\right\rbrace }

\title{Deterministic and stochastic damage detection via dynamic response analysis}

\author{Michael Oberguggenberger\thanks{Unit of Engineering Mathematics, University of Innsbruck,
Technikerstra\ss e 13, 6020 Innsbruck,
Austria (michael.oberguggenberger@uibk.ac.at)}
\and
Martin Schwarz\thanks{Unit of Engineering Mathematics, University of Innsbruck,
Technikerstra\ss e 13, 6020 Innsbruck,
Austria
(martin.schwarz@uibk.ac.at)}
}

\begin{document}
\maketitle

\begin{abstract}
The paper proposes a method of damage detection in elastic materials, which is based on analyzing the time-dependent (dynamic) response of the material excited by an acoustic signal. A case study is presented consisting of experimental measurements and their mathematical analysis. The decisive parameters
(wave speed and damping coefficient) of a mathematical model of the acoustic wave are calibrated by comparing the measurement data with the numerically evaluated exact solution predicted by the mathematical model. The calibration is done both deterministically by minimizing the square error over time and stochastically by a Bayesian approach, implemented through the Metropolis-Hastings algorithm. The resulting posterior distribution of the parameters can be used to construct a Bayesian test for damage.
\end{abstract}

\section{Introduction}
\label{sec:intro}

Using elastic waves in solids for identifying material properties has a long history going back to the 19th century with the pioneering work of Rayleigh, Lamb and Love \cite{Achenbach1976}. In the second half of the 20th century, wave propagation analysis became a central field in seismology \cite{Bleistein2001} and in the material sciences \cite{ewins1984modal}. Especially the analysis of ultrasonic waves, as a part of vibration based condition monitoring \cite{Carden2004}, has become a prominent field in structural health monitoring.

Specifically with respect to structural health monitoring, changes in material properties, such as damage or fatigue of a structure, are sought to being detected by means of the response of the structure to an excitation by an acoustic wave. The focus of this paper will be on acoustic waves induced by a piezoelectric transducer, and measured at the same position or other sensor locations.

In the literature, a wealth of methods have been developed, among them: (a) modal analysis \cite{ewins1984modal}; (b) frequency based methods \cite{Lakshmanan2010,Ostachowicz2019}; (c) general spectral methods \cite{brandt2011noise,Palacz2018}; (d) comparison with finite element solutions \cite{Krawchuk2003,gopalakrishnan2007spectral,Kumar2017,Lammering2018}; (d) Rayleigh, Lamb and Love waves \cite{Lammering2018}; (e) nonlinear analyses \cite{Kerschen2006}; (f) guided waves \cite{Zhao2019}; (g) wavelet analysis \cite{Tian2003}; (h) time series methods \cite{Nair2006,Ostermann2016}; (k); statistical indicators \cite{Zhao2008}; artificial intelligence \cite{Carden2004,Vitola2016}, just to cite a few representative papers. For further information, we refer to the monographs \cite{brandt2011noise,Fahy2007,Krautkraemer1990,Stepinski2013}, the classical survey \cite{Doebling1998} and the more recent survey papers \cite{Carden2004,Palacz2018,Deraemaeker2010,Ostachowicz2010,Fan2011,Marcantonio2019}.

Ultrasonic imaging can be handled in one-, two- and three-dimensional elasticity. Specific lower dimensional applications are to damaged rods \cite{Kumar2017,Krawchuk2006a,Krawchuk2006b} as well as two-dimensional composites \cite{Zhao2019,Ullah2011}, including damage location using sensor configurations \cite{Wandowski2011}. In industrial applications, it is common practice to measure the transit time of the reflected pressure wave, induced by a piezoelectric transducer, and to move the transducer from point to point along the structure. This amounts to measuring the propagation speed of an approximately one-dimensional signal.

Especially in seismology and geotechnics \cite{Manolis2002,Fouque2007}, the random structure of the soil has been incorporated in the study of wave propagation in more recent years. The same applies to material science, see \cite{Demmie2016} and references therein.

The present paper is part of a research project addressing linear wave propagation in random media. The main features of our approach are: (a) models are set up in a way so that exact solutions can be used and (b) stochastic analyses are employed to obtain indicators for changes in material parameters and for damage detection. The program encompasses unidirectional propagation (transport equations), one-dimensional acoustic waves and wave propagation in three-dimensional linearly elastic solids. In the present paper, we present a theoretical and experimental analysis of ultrasonic inspection of composite plates. The acoustic wave produced by a piezoelectric transducer is modelled as a one-dimensional wave issuing from the (moving) location of the transducer, which also acts as recording sensor for the reflected wave. The corresponding initial-boundary value problem for the damped wave equation is solved explicitly, using Fourier transform techniques. The propagation speed and damping coefficient are directly related to elastic parameters of the material and can be calibrated by comparison with the time-dependent signal. In addition, a stochastic, Bayesian parameter estimation allows one to design hypothesis tests for critical thresholds of the model parameters. A three-dimensional analysis by means of Fourier integral operators will be presented in a forthcoming paper.

In the specific case of ultrasonic pulse echo measurements, the presented work includes an experimental part of performing the measurements, and a theoretical part of setting up the wave propagation model, calibrating the parameters, finding the posterior distributions of the calibrated parameters, and setting up Bayesian hypothesis tests for damage.

The experimental set-up included four carbon fiber composite plates, three of which were damaged by a localized impact.
The plates were measured with a piezoelectric transducer in impulse echo mode.
As mentioned, the wave propagation through the material at a single measuring location is modeled by means of a one-dimensional wave equation with a damping term included. This plane wave assumption is justified through the respective dimensions of the transducer head and the thickness of the plates. Having set up the wave propagation model, including the incoming and reflecting boundary conditions, the two parameters (wave speed and damping coefficient) are calibrated by fitting the numerically evaluated exact solution to the measurements.

The calibration is first performed deterministically at each location by minimizing the mean square error over the observation time. Repeating this procedure on a grid of points on the plate allows one to identify locations in which the parameters deviate from the overall mean values.

However, a more stringent approach to determining the statistical properties of the parameter values at single locations is stochastic parameter calibration by means of Bayesian methods. Indeed, based on a priori bounds on the parameters and the likelihood function (given by the probability distribution of the error between certain measured and computed features of the response), the posterior (joint) distribution of the wave speed and the damping coefficient can be obtained at each location. What is more, the posterior density admits to determine credible regions, by means of which Bayesian hypothesis tests can be designed. One may rephrase the null hypothesis of undamaged material as a non-critical domain for the model parameters; the posterior probability of the null hypothesis is nothing but the posterior probability of the non-critical domain. The so designed Bayesian test, performed at a 1\% rejection threshold, enables the location of damaged points on the plate. The results are in accordance with the deterministic approach, but the Bayesian test contains more statistical information.

The outline of this paper is as follows. In the second section the measurement set-up will be described in more detail.

The third section is devoted to setting up the mathematical model of the waves traveling through the plate. We construct a Fourier transform based solution operator and we show that the model is capable of reproducing the measured signals.

In the fourth section we describe methods for parameter estimation and testing. In the first part we deterministically calibrate the wave speed and the damping coefficient to get a best fit with the measured signal at each location of the plate, using minimization of the mean square error between measured signal and computed solution over time. In this way, damage can be localized by observing deviations of the parameters from their nominal values. In the second part we apply Bayesian methods to compute the posterior distribution of the parameters. To this end, we implement the Metropolis-Hastings algorithm, generating a Markov chain the stationary distribution of which is the posterior distribution. Thereby, the uncertainty of the parameter estimation for each single location on the plate can be quantified. Finally, in the third part, we construct a Bayesian test for damage, which allows one to identify damaged areas on the plate. The paper concludes with a discussion and a summary.

The paper is based on the work \cite{Schwarz:Thesis:2019} of the second author. Partial results on transport equations and on acoustic waves in three-dimensional elasticity have been reported in \cite{MOMS:14, Baumgartner2017, LOS:17, MOMS2018}.

\section{Ultrasonic impulse echo}
\label{sec:ultra}

\begin{figure}[tbh]
	\centering
	(a)\quad \includegraphics[width=0.4\linewidth]{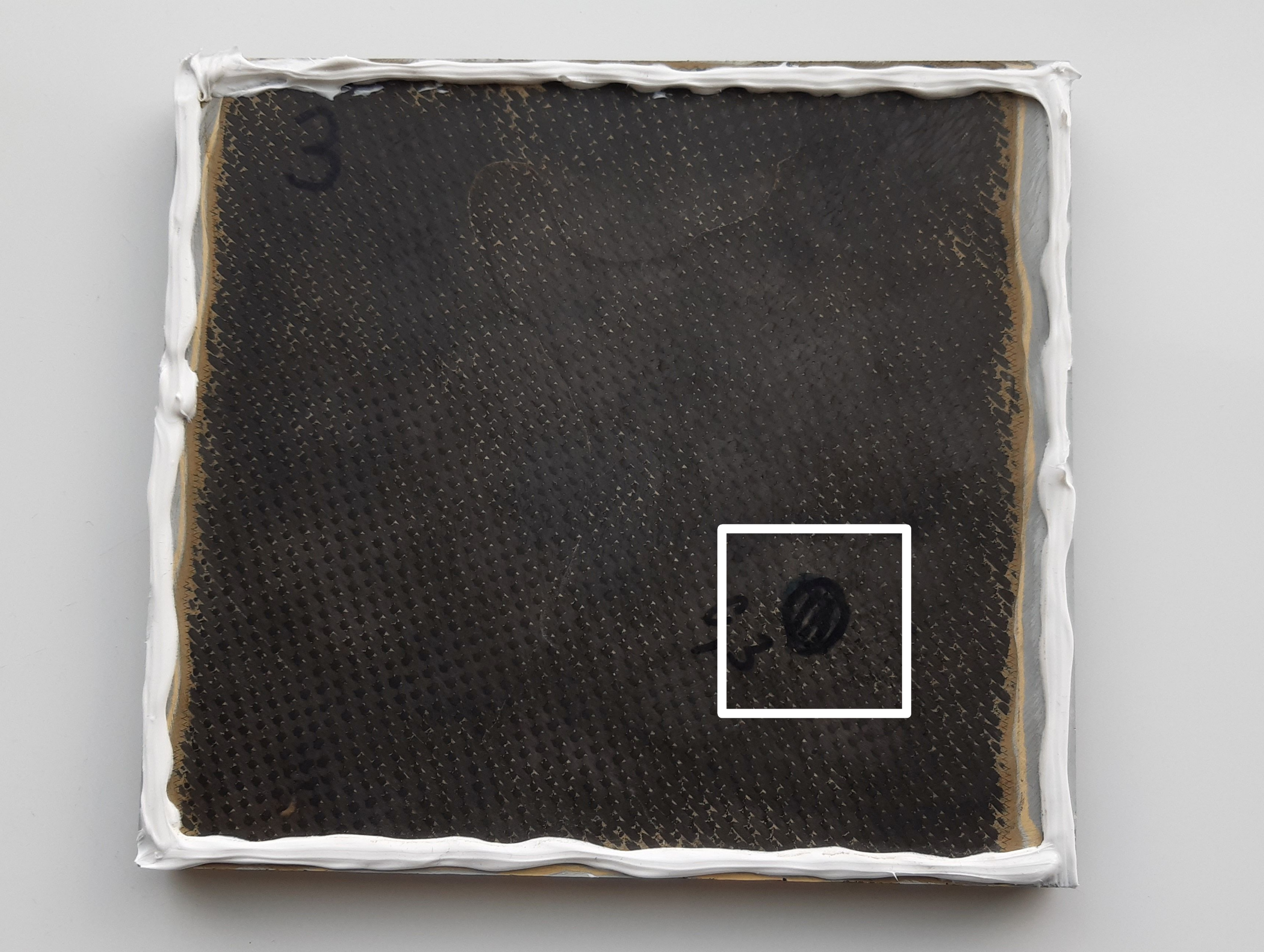} \quad
    (b)\quad \includegraphics[width=0.4\linewidth]{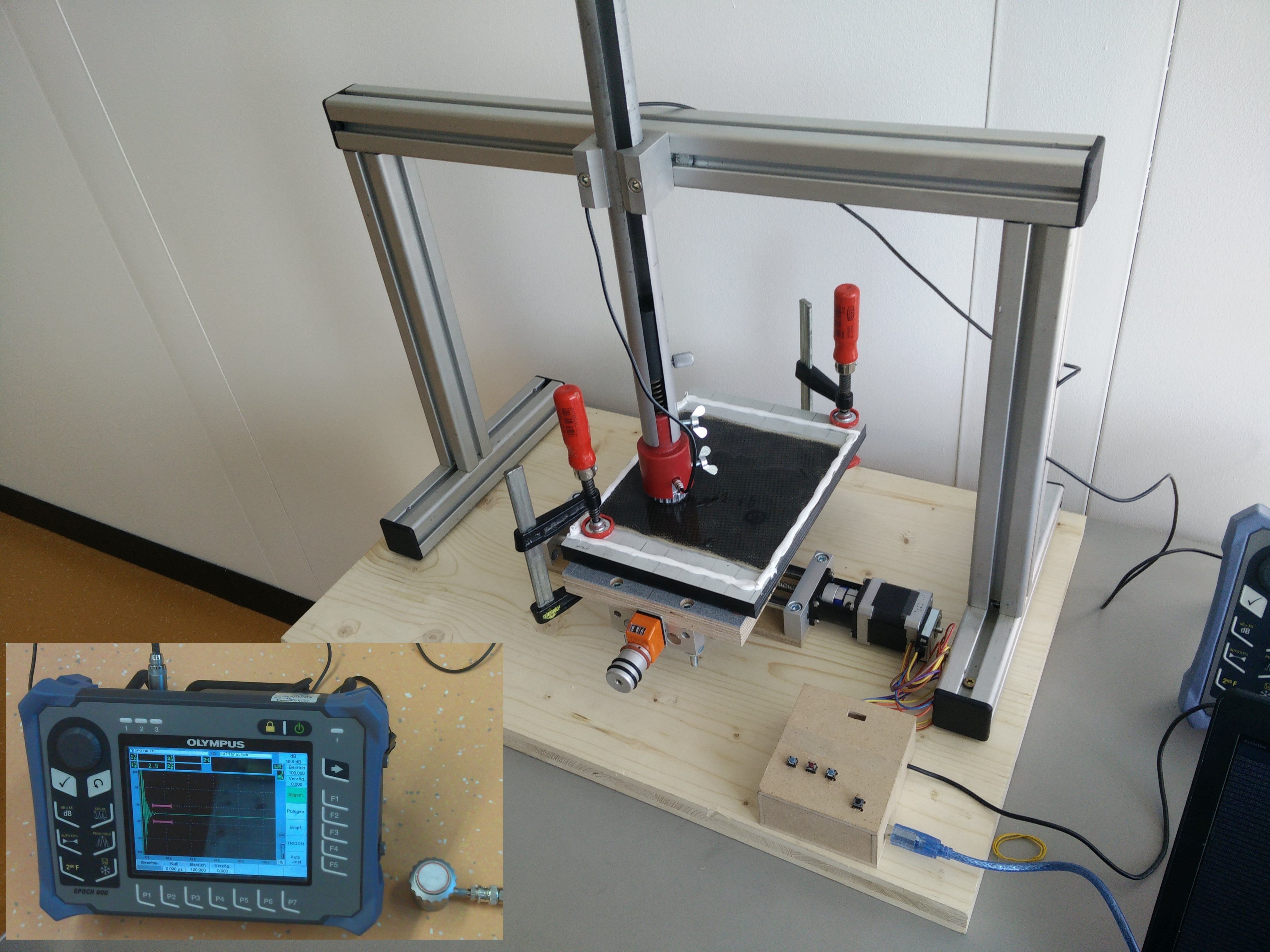}
	\caption[Experiment]{Experimental setup: (a) damaged plate specimen, impact can be seen as dark spot inside white square in lower right quadrant; (b) measuring platform, picture of ultrasonic detector inserted}
	\label{fig:experiment}
\end{figure}

As mentioned in the introduction, the measurement data were obtained by recording the response of four carbon fiber composite plates to an ultrasonic signal. Three of the specimen had been damaged by a high speed impact and the fourth plate was undamaged. The dimensions of the specimen were 105 mm times 95 mm with a thickness of about 13 mm. (The exact thickness is not relevant for the analysis.)
All four plates were scanned with a horizontal resolution of $5\times5$ millimeters. In addition, the damaged area was scanned with a horizontal resolution of $1\times1$ mm. The damaged specimen analyzed further in the paper is shown in Figure \ref{fig:experiment}(a).

The principle of an ultrasonic impulse echo measurement is the following: A piezoelectric transducer produces an ultrasonic pulse at the surface of the plate which then goes through the plate as an elastic wave. The wave is reflected at the bottom and comes back to the top. The transducer measures the amplitude at the top over time, which is recorded as voltage by means of an oscilloscope. In order to have good contact between transducer and plate a water film was placed on top of the plate. As pulse generating and recording device an Olympus${}^{\mbox{\footnotesize\textregistered}}$  Epoch 600 Ultrasonic Flaw Detector was used. The pulser voltage was set to 400 V; a transducer of diameter 22 mm and resonance frequency of 1 MHz was employed. The oscilloscope had a resolution of $400$ MHz in time. For each scan, the recorded period was $35$ $\mathrm{\mu}$s and thus $14000$ data points per scan were acquired. The amplitude resolution of the oscilloscope was $512$ points, where number $256$ represented the zero line.
For the subsequent mathematical analysis, the signal was normalized by its maximal value of $256$ points. Accordingly, the analyzed signal was dimensionless with a range between $-1$ and $1$. In accordance with widespread usage (e.g., \cite[Chapter 10]{Lammering2018}, \cite{Zhao2019}), \cite[Chapter 5]{Stepinski2013}), we shall refer to it as \emph{normalized amplitude}.

The transmission from the oscilloscope to the PC sometimes produced scrambled signals. To detect such measurement errors, the transmitted signal was analyzed by an automated script. A faulty signal typically had a jump at a certain time point. Thus, the successive differences in time were considered. During the excitation period larger differences were accepted, whereas in the arriving echoes the differences had to be small, since the signal should be continuous. Measurements of signals classified as erroneous were repeated.

\begin{figure}[tbh]
	\centering
	\raisebox{6mm}{(a)}\quad \includegraphics[width=0.8\linewidth]{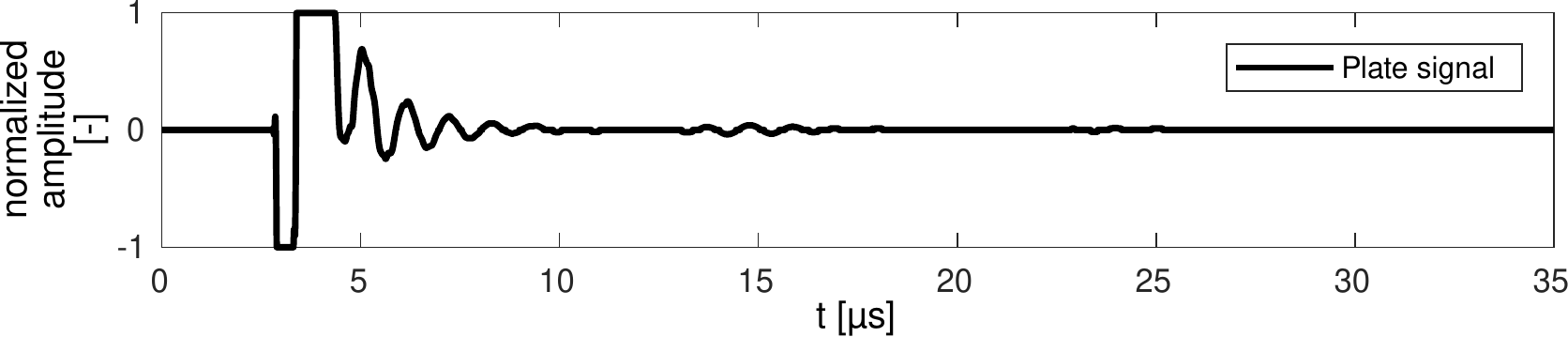}\\
    \raisebox{6mm}{(b)}\quad \includegraphics[width=0.8\linewidth]{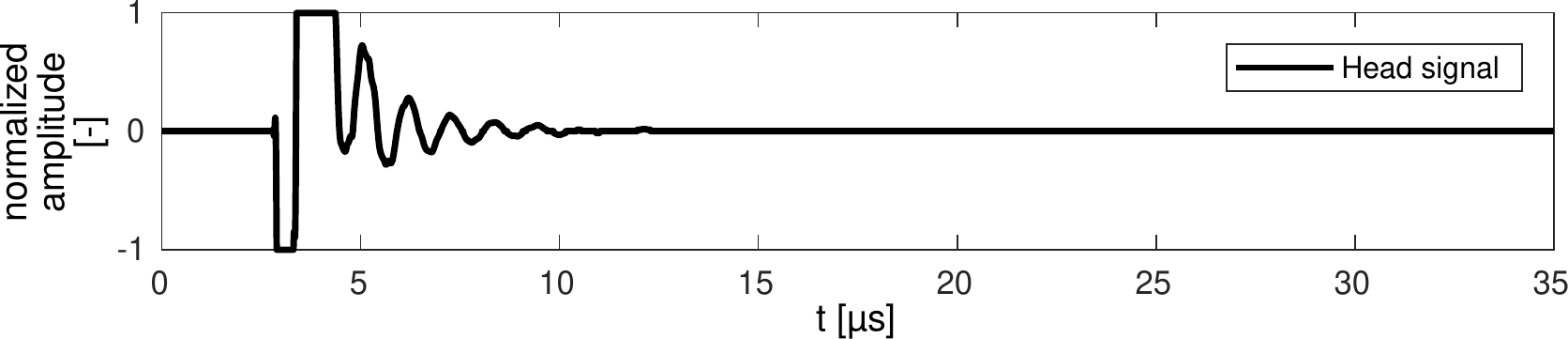}\\
    \raisebox{6mm}{(c)}\quad \includegraphics[width=0.8\linewidth]{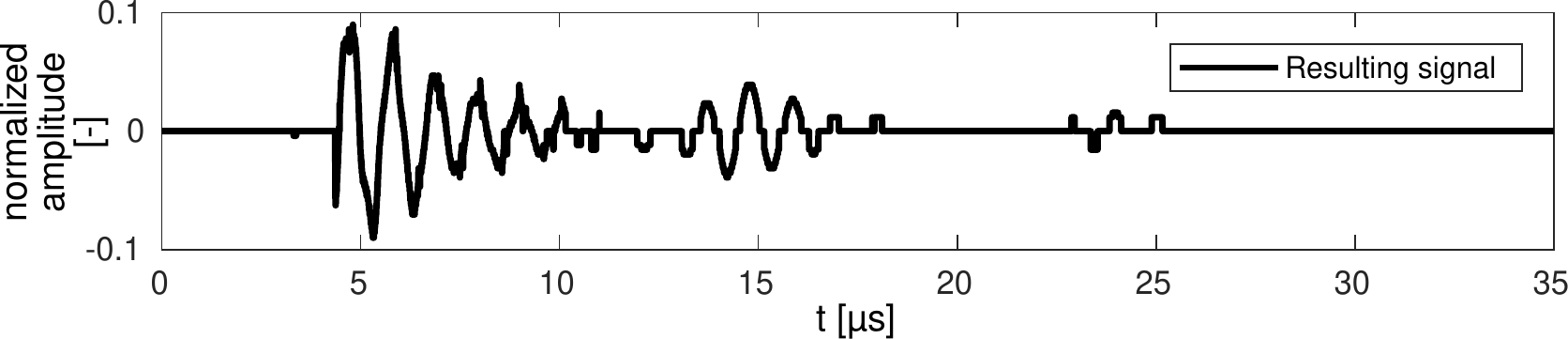}
	\caption[Transducer Head]{Removing the self-referential part of the transducer head's signal: (a) plate signal; (b) free head signal; (c) adjusted
     signal obtained by subtraction}
	\label{fig:signalminuskopf}
\end{figure}

Furthermore, the signal of the transducer not contacting the plate was subtracted from the signal touching the plate, since the transducer head measured the vibrations within itself, too. The resulting adjusted signal depicts the actual vibrations in the plate (see Figure \ref{fig:signalminuskopf}).

As a final remark we would like to point out that the excitation is done by sending an electrical rectangular pulse to the transducer head. This pulse causes the transducer head to generate a sinusoidal vibration. However, this pulse is stronger than the oscilloscope can measure, which produced an overflow in amplitude direction for the first 4 $\mu$s. This means that the initial pulse cannot be measured exactly and by subtracting the signals it is set to zero (although it is nonzero). As a consequence, the first period of the initial pulse is measured as zero. Therefore, the echoes have one more period than the initial pulse in the measurements, as can be seen in Figure \ref{fig:simvsmeasurement} (located on the time axis between $12$ $\mu s$ and $13$ $\mu s$).

\section{Mathematical modeling: 1D telegraph equation}
\label{sec:mathmod}

\begin{figure}[htb]
	\begin{center}
		\raisebox{4mm}{(a)}\quad\includegraphics[width=0.8\linewidth]{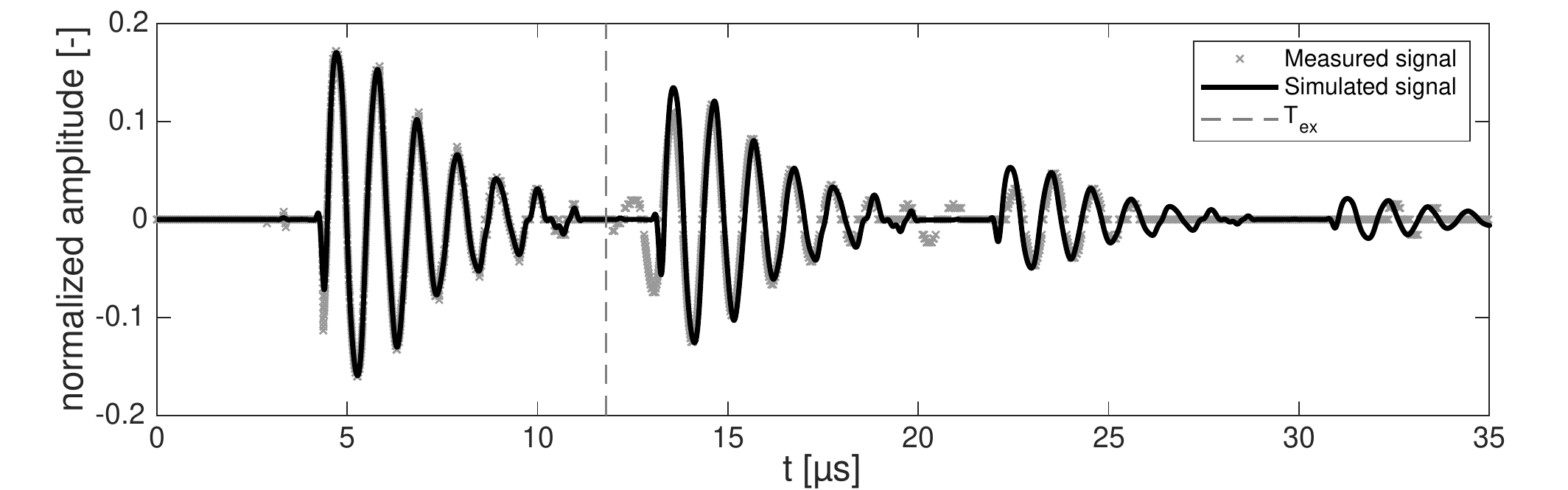}\\
		\raisebox{4mm}{(b)}\quad\includegraphics[width=0.8\linewidth]{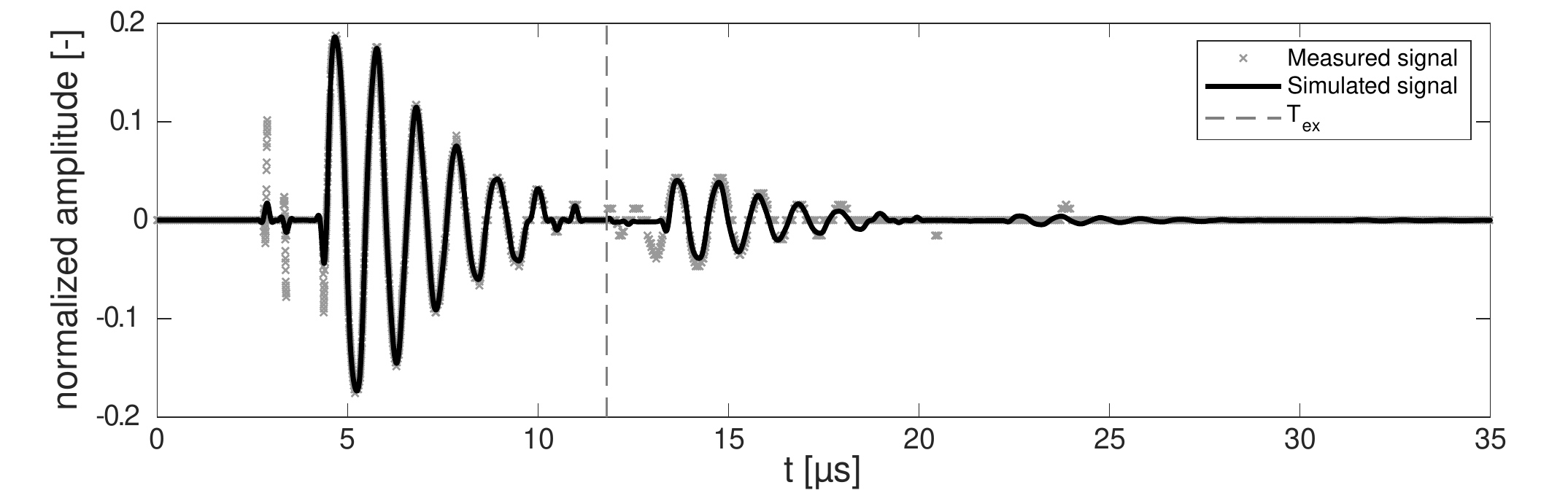}
	\end{center}
	\caption{Measured signal (black curves) and signal computed with calibrated parameters $b$ and $c$ (gray curves) at: (a) an undamaged location and (b) a damaged location; $T_{\text{ex}} = 11.8\; \mu$s is indicated by a dotted line}
	\label{fig:simvsmeasurement}
\end{figure}

To set up a simple mathematical model that nevertheless captures the features of interest, we assume that the undamaged plate is a linearly elastic, homogeneous, isotropic continuum of constant density $\rho$. The in-plane coordinates will be denoted by $x$ and $y$, the vertical direction by $z$ with $z=0$ corresponding to the upper surface of the plate.
Further, we assume that the transducer induces a plane wave moving in $z$-direction of the plate. This can be justified, since the diameter of the transducer is larger than the thickness of the plate (approximately $2:1$). Plane waves can be reduced to the one-dimensional acoustic wave equation
\[
  \partial_{tt} u(z,t) - c^2\, \partial_{zz}u(z,t)=0
\]
for the displacement $u(z,t)$ in $z$-direction at time $t$. The wave speed is related to the Lam\'e constants $\lambda$, $\mu$ of the material through
\begin{equation}\label{eq:Lame}
c^2 = (\lambda + 2\mu)/\rho,
\end{equation}
see \cite[Sections 1.2 and 1.3]{Achenbach1976}. The acoustic wave equation neglects attenuation which is due to absorption and to scattering of waves at inhomogeneities of the material \cite[Chapter 6]{Krautkraemer1990}.
The attenuation can also be observed in the experimental data, see Figure \ref{fig:simvsmeasurement}(a). Thus it is justified to add a damping term to the wave equation, resulting in the telegraph equation. The bottom side of the plate is assumed to be stress free, which implies that the $z$-derivative of $u$ vanishes.

During the time interval $[0, T_{\text{ex}}]$ the excitation due to the piezoelectric transducer produces a displacement $f(t)$ at the top side of the plate \cite{Fahy2007}. We assume $f$ to be smooth and to be of compact support in $(0,T_\text{ex})$. During the observation period $(T_{\text{ex}},T_{\text{end}}]$, a stress free boundary on top side is assumed, and the time-dependent displacement induces the signal recorded by the transducer.

As noted in Section \ref{sec:ultra}, the exact value of the plate thickness is irrelevant to the analysis. (One-sided ultrasonic measurements do not allow one to measure the absolute values of the plate thickness and the wave speed simultaneously.) We assume from now on that the thickness of the plate has the constant value $L$, which also serves as our (virtual) length unit.

This results in the following equations, valid for $0\leq z \leq L$:
\begin{subequations} %\label{eqn:1d_tel_eqns}
	\begin{align} \label{eqn:1du}
	u(z,t)=\begin{cases}
	v(z,t)& \text{ if }t\in[0,T_{\text{ex}}]\\
	w(z,t)& \text{ if }t\in (T_{\text{ex}},T_{\text{end}}],
	\end{cases}
	\end{align}
	where
	\begin{align}\label{eqn:1dv}
	\left\{\begin{aligned}
	\partial_{tt} v(z,t)+b\, \partial_t v(z,t) - c^2\, \partial_{zz}v(z,t)&=0\\
	v(z,0) &= 0\qquad & \partial_t v(z,0) &= 0\\
	v(0,t) &= f(t)& \partial_z v(L,t)&= 0
	\end{aligned}\right. \end{align}
	and
	\begin{align} \label{eqn:1dw}
	\left\{\begin{aligned}
	\partial_{tt} w(z,t)+b\, \partial_t w(z,t) - c^2\, \partial_{zz} w(z,t)&=0\\
	w(z,T_{\text{ex}}) - v(z,T_{\text{ex}})&=0 & \partial_t w(z,T_{\text{ex}}) -\partial_t  v(z,T_{\text{ex}})&=0\\
	\partial_z w(0,t) &= 0& \partial_z w(L,t)&= 0 .
	\end{aligned}\right.
	\end{align}
\end{subequations}

In this set-up, $f(t)$ for $t<T_{\text{ex}}$ and $w(0,t)$ for $t>T_{\text{ex}}$ correspond to the readings of the oscilloscope and couple the measurements with the solution to \eqref{eqn:1du}--\eqref{eqn:1dw}. The output of the oscilloscope is voltage. As mentioned in the introduction, the signal is normalized by its maximally measurable value. We refer to this dimensionless quantity as \emph{normalized amplitude}.

The graphs in Figure \ref{fig:simvsmeasurement} confirm the good coherence between the measured signal and the solution of \eqref{eqn:1du}--\eqref{eqn:1dw} at $z=0$, the top of the plate. The parameters $b$ and $c$ were calibrated such that the mean square error of the computed and the measured signal was minimal. The calibration procedure will be explained in detail in the next section. As mentioned before, the measured echoes have one more period in the time between $12$ $\mu s$ and $13$ $\mu s$ than the computed ones. This is due to the fact that the force term cannot be measured in the first 4 $\mu s$ and is set to zero.

Various analytical methods for solving equations \eqref{eqn:1dv} and \eqref{eqn:1dw} are available (see e.g. \cite{Graff:75}). It turned out that for numerical reasons, problem \eqref{eqn:1dv} is advantageously solved by applying Fourier transform in time direction, which can be evaluated very quickly. Problem \eqref{eqn:1dw} is more easily solved by Fourier series expansion.

We start with solving \eqref{eqn:1dv}. In order to apply the Fourier transform in the time direction, we first have to extend the time domain to the full time axis. We define
\[ F(t)=\begin{cases}
f(t) & t\in(0,T_{\text{ex}})\\
0 & \text{else}.
\end{cases} \]
At this stage, we let $v$ be the solution of \eqref{eqn:1dv} for $t\in[0,\infty)$ (not only $t\in[0,T_\text{ex}]$). We further extend $v$ to negative times by setting
\begin{align*}
V(z,t)=\begin{cases}
v(z,t)& t \geq 0\\
0 & t < 0.
\end{cases}
\end{align*}
If $V$ is at least twice continuously differentiable, it satisfies
\begin{align}\label{eqn:1dvv}
\left\{\begin{aligned}
\partial_{tt} V(z,t)+b\, \partial_t V(z,t) - c^2\, \partial_{zz}V(z,t)&=0\\
V(0,t) &= F(t)& \partial_z V(L,t)&= 0\\
V(z,\cdot)\big|_{t<0} &\equiv 0.
\end{aligned}\right. \end{align}

We formally deduce the solution: If we apply the Fourier transform in time direction (denoted as $\nF_{t\to\tau}[V]=\widetilde{V}$), we get the ordinary differential equation
\begin{align*}
\left\{\begin{aligned}
-\tau^2\widetilde{V}(z,\tau)+i \tau b\widetilde{V}(z,\tau) - c^2\, \partial_{zz}\widetilde{V}(z,\tau)&=0\\
\widetilde V(0,\tau) =\widetilde{f}(\tau) \qquad  (\partial_z \widetilde V)(L,\tau) &= 0.\\
\end{aligned}\right.
\end{align*}
This is solved by
\begin{align*}
\widetilde{V}(z,\tau)= C_1 (\tau) \re ^{- B(\tau) z} + C_2 (\tau) \re^{ B(\tau) z} ,
\end{align*}
where $B(\tau)=\frac{1}{c} \sqrt{-\tau^2+ib\tau}$, and $\sqrt{\cdot}$ is the principle branch of the complex root, and $\sqrt{0}=0$. The constants $C_1$ and $C_2$ are given by
\begin{align*}
C_1(\tau)=\widetilde{F}(\tau) \frac{\re^{ B(\tau) L}}{\re^{B(\tau)L}+\re^{-B(\tau)L}} \qquad \text{and} \qquad C_2(\tau)=\widetilde{F}(\tau) \frac{\re^{- B(\tau) L}}{\re^{B(\tau)L}+\re^{-B(\tau)L} }
\end{align*}
and consequently
\[ \widetilde{V}(z,\tau)=\widetilde{F}(\tau) \frac{\re^{(L-z) B(\tau)}+\re^{-(L-z) B(\tau)}}{\re^{L B(\tau)}+\re^{-L B(\tau)}}. \]
Applying the inverse Fourier transform yields the formal solution
\begin{align}\label{eqn:V_from_Fourier_transform}
V(z,t) = \int_{-\infty}^{\infty} \re^{i\tau t} \widetilde{V}(z,\tau) \dbar \tau
\end{align}
where $\dbar \tau$ is shorthand for $\d \tau/2\pi$.\\

\noindent
{\bf Proposition.}
\emph{Let $V$ be as above. Then \eqref{eqn:V_from_Fourier_transform} is a convergent integral and {$V$ is infinitely differentiable} and satisfies \eqref{eqn:1dvv}.}

\begin{proof}
	We will only sketch how to prove this. The steps are as follows:
    \begin{itemize}
    	\item Since $F$ is a Schwartz function (infinitely differentiable, faster decay than any negative power of $\abs{t}$), the time Fourier transform of $F$ is a Schwartz function, too. Thus, it suffices to show that \[ \frac{\re^{(L-z) B(\tau)}+\re^{-(L-z) B(\tau)}}{\re^{L B(\tau)}+\re^{-L B(\tau)}} \]
    	is bounded in $\tau$. Then \eqref{eqn:V_from_Fourier_transform} is a convergent integral.
    	\item To show that $V$ is smooth one needs to show that differentiation under the integral sign is justified.
    	\item It is trivial to show that \eqref{eqn:V_from_Fourier_transform} satisfies the boundary condition. Finally, we use the Paley-Wiener theorem, which gives a sufficient condition on $\widetilde{V}$ such that $V(z,\cdot)\big|_{t<0}\equiv 0$.
	\end{itemize}
	
	A detailed proof of all steps can be found in \cite{Schwarz:Thesis:2019}.
\end{proof}

\noindent We can compute initial values of \eqref{eqn:1dw} by
\[ w(z,T_{\text{ex}}) = \int_{-\infty}^{\infty} \re^{i\tau T_{\text{ex}}} \widetilde{V}(z,\tau) \dbar \tau  \quad \text{and} \quad (\partial_t w)(z,T_{\text{ex}}) = \int_{-\infty}^{\infty} \re^{i\tau T_{\text{ex}}} i\tau  \widetilde{V}(z,\tau) \dbar \tau.\]

Both the Fourier transform and the inverse Fourier transform can be numerically approximated by the \emph{fft} respective \emph{ifft} algorithm. Since the force term was smooth enough, the transformed signal decayed fast enough to get a good numerical approximation.

For the numerical calculation we set $L=1$. (As noted at the beginning of Section\;\ref{sec:mathmod}, this is no restriction.) As spatial discretization we chose $\Delta z=0.001$. The time discretization was the same as from the oscilloscope, i.e. $T_{\text{max}}=35\; \mu$s and $\Delta t=0.0025\; \mu$s. Furthermore, we set $T_{\text{ex}}=11.8\; \mu$s. This is the time after which the force term $f$ was zero in all measurements.

In order to have a smooth input signal $f$, the measured signal of the transducer was regularized by smoothing out the high frequencies. For this purpose we multiplied $\widetilde f$ with a Tukey window (see e.g. \cite{Harris1978}), removing frequencies beyond 6.5 MHz.

We use the reflection principle (see e.g. \cite[Chapter 2, Section 2]{Graff:75}) to extend \eqref{eqn:1dw} to the strip $[-L,L]\times [T_{\text{ex}},\infty)$ as
\begin{align*}
W(z,t)=\begin{cases}
w(z,t) & z\geq 0\\
w(-z,t) & z < 0
\end{cases}
\end{align*}
and
\begin{align*}
{V}_e(z,T_{\text{ex}})=\begin{cases}
{V}(z,T_{\text{ex}}) & z\geq 0\\
{V}(-z,T_{\text{ex}}) & z <0.
\end{cases}
\end{align*}
Then $W$ satisfies
\[ \dgl{\p_{tt}W(z,t)+b\p_tW(z,t) -c^2 \p_{zz}W(z,t)&=0\\
	W(z,T_{\text{ex}})- {V}_e(z,T_{\text{ex}})&=0 \qquad \p_tW(z,T_{\text{ex}})-\p_t {V}_e(z,T_{\text{ex}})=0\\
	\p_z W(-L,t) &=0 \qquad \p_z W(L,t)=0.}\]

Since $W$ is even and continuously differentiable, the homogeneous Neumann boundary condition is equivalent to periodic boundary conditions. This can be shown by iteratively applying the reflection principle and extending $W$ to the whole space by
\[ {W}_e(z,t)= \begin{cases}
\qquad\vdots& \qquad \vdots\\
W(z-2L,t) & \text{if } z\in [-3L, -L)\\
W(z,t) & \text{if } z\in[ -L, L)\\
W(z-2L,t) & \text{if } z\in[ L, 3L).\\
\qquad\vdots& \qquad \vdots
\end{cases} \]
Then ${W}_e$ is $2L$-periodic. On the other hand, if $W$ is even and periodic, then
$W(z+L)=W(z-L)=W(-z+L)=W(-z-L)$, and thus $W_z(L)=W_z(-L)=0$.

Thus, we can equivalently solve the periodic boundary value problem
\[ \dgl{\p_{tt}W(z,t)+b\p_tW(z,t) -c^2 \p_{zz}W(z,t)&=0\\
	W(z,T_{\text{ex}})-{V}_e(z,T_{\text{ex}})&=0 \qquad \p_tW(z,T_{\text{ex}})-\p_t{V}_e(z,T_{\text{ex}})=0\\
	W(-L,t)-W(L,t)&=0 \qquad \p_zW(-L,t)-\p_zW(L,t)=0.}\]
We make the ansatz $W(z,t)=\sum_{k\in\mathbb{Z}} a_k(t) \re^{i{k\pi}{}z/L}$, which is $2L$-periodic in space. Plugging it into the equation yields
\[ a_k(t)=A_k \ \exp\bigg(-\frac{b}{2}t+i\sqrt{\frac{c^2 k^2 \pi^2}{L^2}-\frac{b^2}{4}}t\bigg)+ B_k \ \exp\bigg(-\frac{b}{2}t-i\sqrt{\frac{c^2 k^2 \pi^2}{L^2}-\frac{b^2}{4}}t\bigg).\]
The constants $A_k$ and $B_k$ are determined from a Fourier series expansion of ${V}_e(z,T_{\text{ex}})$ and $\p_t {V}_e(z,T_{\text{ex}})$.

For the numerical calculation we used the same grid size: $L=1, \Delta z=0.001.$ The time discretization was the same as from the oscilloscope, i.e. $T_{\text{max}}=35$ $\mu$s and $\Delta t=0.0025$ $\mu$s. The computed signals at an undamaged and a damaged location are shown in Figure \ref{fig:simvsmeasurement}. The evaluation of the numerical solution takes approximately 2.8 seconds on a PC.

\section{Parameter estimation and tests}

The mathematical model described above will serve as an input-output model. As input we have the measured excitation $f$ and the (unknown) material parameters $b$ and $c$. As output we get $g_{\text{comp}}(t)=u(0,t)$. Furthermore, we will call $g_{\text{meas}}(t)$ the measured signal.

In the following subsection we will calibrate the material parameters $b$ and $c$ at each measurement location. In the second subsection we will estimate the posterior distribution of $b$ and $c$ for every measurement location. According to our length convention, the dimension of $c$ is L/$\mu$s. The dimension of $b$ is 1/$\mu$s.

\subsection{Parameter estimation via optimization} \label{subsec:optim}

\begin{figure}[tbh]
	\centering
	\raisebox{6mm}{(a)} \includegraphics[width=0.4\linewidth]{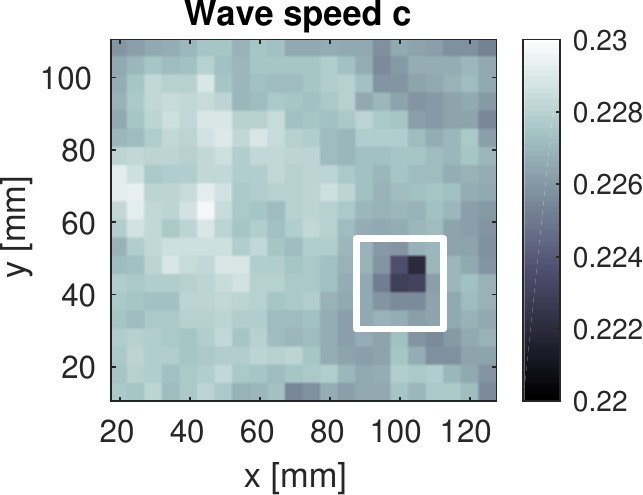} \qquad
    \raisebox{6mm}{(b)} \includegraphics[width=0.38\linewidth]{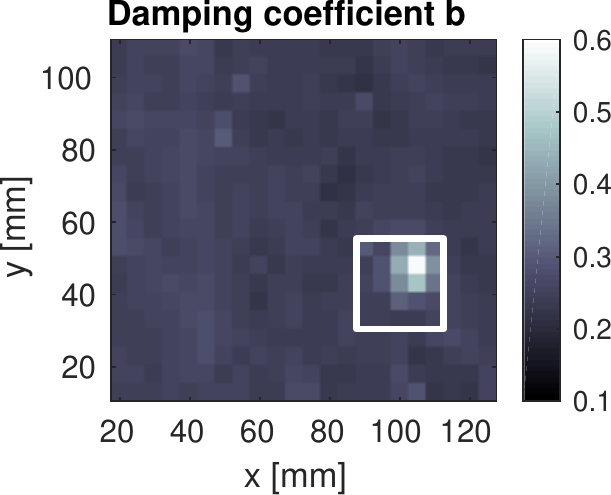}\\[6pt]
   \hspace{7pt} \raisebox{6mm}{(c)} \includegraphics[width=0.4\linewidth]{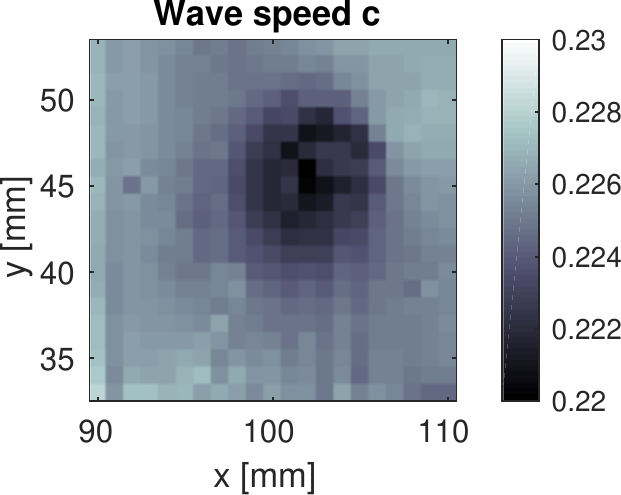} \qquad
    \raisebox{6mm}{(d)} \includegraphics[width=0.38\linewidth]{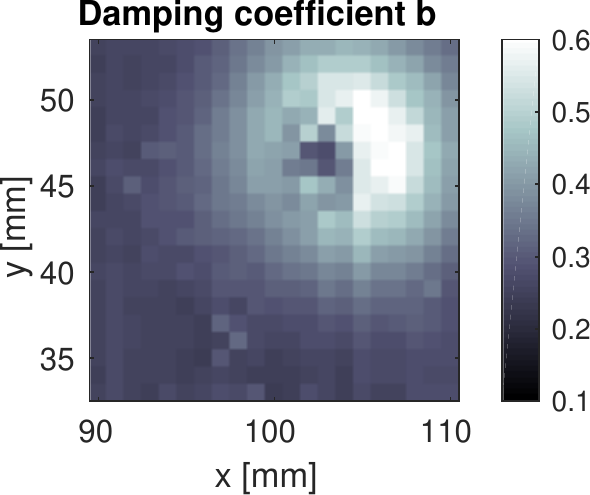}
	\caption{Spatial resolution of optimized parameters: (a) overall view of wave speed $c$; (b) overall view of damping coefficient $b$;
    (c) and (d) high resolution of damaged region. The gray scale is in units L/$\mu$s for $c$ and in units 1/$\mu$s for $b$. The $x$- and $y$-coordinates refer to the location on the positioning platform.}
	\label{fig:platte3}
\end{figure}

In order to estimate $b$ and $c$ in a given location, we solve the non-convex optimization problem
\[ \arg\min_{b,c} \norm{ g_{\text{meas}}-g_{\text{comp}}}_{L_2([0,T_{\text{end}}])}.\]
The optimal parameters were computed using the \textit{Nelder-Mead algorithm}. Because the problem is nonconvex we had to be careful to choose a good initial value. As initial value, we chose the pair $(b,c)=(0.2,0.22)$, which seemed to work well for most locations. Figure \ref{fig:simvsmeasurement} shows the comparison between the measured data and the computed signal with optimized parameters at a single location. The spatial resolution of the optimized parameters $b$ and $c$ of the plate specimen under investigation is displayed in Figure \ref{fig:platte3}.

We observe that the wave speed decreases and the damping coefficient increases in the damaged region. The damage affects the stiffness of the material, which makes waves travel more slowly. Indeed, in an elastic medium the Lam\'e constants and hence the square of the wave speed are proportional to the modulus of elasticity $E$, compare Equation \eqref{eq:Lame} and \cite[Table 2.1]{Achenbach1976}, and damage generally goes together with a decrease of $E$ \cite[Section 2.1.2]{Murakami2012}, \cite{Alves2000}. On the other hand, one expects delamination within the damaged region, which leads to higher damping, that is, attenuation of the signal \cite{Stone1975, Karabutov2014}.

\subsection{Bayesian parameter estimation}
The least square error optimization described above produces a point estimator for the parameters $b$ and $c$. The Bayesian inference described in the sequel provides more information about the estimated parameters. Using prior knowledge of the parameters this method gives a probability distribution of the parameters. This can be used to get an estimator for $b$ and $c$ but also to assess the variation of the parameters.

For this reason consider a state space $X\subset\RR^n$ and a parameter space $\Theta\subset \RR^m$. Let $\bx$ resp. $\btheta$ be random variables with values in $X$ resp. $\Theta$ and marginal density $f_{\bx}$ resp. $f_{\btheta}$. Then, by Bayes' theorem
\begin{equation} \label{eqn:cont_prob}
\PP({\btheta}\in B|{\bx}=\balpha)=\frac{\int_{B} f_{{\bx}|{\btheta=\bbeta}}(\balpha) f_{\btheta}(\bbeta) \d \bbeta}{\int_{\Theta} f_{{\bx}|{\btheta}=\bbeta}(\balpha) f_{\btheta}(\bbeta) \d \bbeta},
\end{equation}
where $B$ is a Borel measurable set and $f_{{\bx}|{\btheta=\bbeta}}$ is the conditional density of $\bx$ given that $\btheta=\bbeta$. In terms of densities one has
\[ f_{{\btheta}|{\bx=\balpha}}(\bbeta)= \frac{f_{{\bx}|{\btheta=\bbeta}} (\balpha) f_{\btheta}(\bbeta) }{f_{\bx}(\balpha)}. \]
We use this rule for Bayesian inference \cite{Ka-Veng2010} with a fixed model: If $\balpha\in X$ is the measured data obtained in an experiment, then the posterior probability density function of the parameters ${\btheta}\in\Theta$ is

$$f_{{\btheta}|{\bx}=\balpha}(\bbeta)= \kappa f_{{\bx}|{\btheta}=\bbeta} (\balpha) f_{\btheta}(\bbeta),$$ where $\kappa$ is a constant depending only on $\balpha$. Here, $f_{\btheta}$  is the prior probability distribution of the parameters ${\btheta}$ and $f_{{\bx}|{\btheta}}$ is the likelihood function.

We assume that $f_{\bx|\btheta}$ is known. In principle, one could compute the posterior distribution directly. However, usually -- and also in our case -- the model is computationally expensive to evaluate, so a direct evaluation of the integrals in \eqref{eqn:cont_prob} is problematic. We will employ the \emph{Metropolis-Hastings algorithm} ({MHA}) instead.

This algorithm allows one to generate a Markov chain with a (up to a normalization constant) given stationary distribution \cite{Young2005}. Some mild assumptions guarantee the convergence of the Markov chain to the stationary distribution. The generated sample can then be used to estimate the conditional probability $f_{\btheta|\bx}$ by various methods, e.g. by applying a smooth kernel density estimator.

For the MHA one needs an initial probability density $p_0$ and a target probability density $p$. Furthermore, one needs a proposal probability density $q(\cdot,\by)$, which may depend on parameter $\by\in\Theta$. Then, the Markov chain is generated as follows.

\begin{itemize}
	\item Sample the first Markov chain link $\bbeta_0$ according to the given initial probability distribution $p_0$.
	\item The $k$th chain link $\bbeta^k$ is generated as follows: Generate a candidate  $\boldeta^k$ according to the proposal distribution $q(\cdot,\bbeta^{k-1})$. Then compute
	\[\pi(\boldsymbol{\eta}^{k},\bbeta^{k-1})=\begin{cases}
	\min\menge{1,\frac{p(\boldeta^{k}) q(\bbeta^{k-1},\boldeta^{k})}{p(\bbeta^{k-1})q(\boldeta^k,\bbeta^{k-1})}} & \text{if }p(\bbeta^{k-1}) q(\bbeta^{k-1},\boldeta^{k}) >0\\
	1 & \text{otherwise}
	\end{cases}\]
	and set $\bbeta^k=\boldeta^k$ with probability $\pi(\boldsymbol{\eta}^{k},\bbeta^{k-1})$. Otherwise set $\bbeta^k=\bbeta^{k-1}$.
	
\end{itemize}

Finally, one arrives at the Markov chain $(\bbeta^0,\bbeta^1,\ldots,\bbeta^N)$ having asymptotically the distribution $p$.

Naturally, the question arises, what criteria guarantee the convergence to the stationary distribution and, secondly, how fast is the convergence.

For the convergence theory we refer to \cite{Schwarz:Thesis:2019,AtchY07,JarnS00,RobeG97,RobeG96}. A possible answer can be summarized as follows:

\begin{itemize}
	\item If the initial density $p_0$ equals the target density $p$, then $\bbeta^k$ is $p$-distributed for $k\in\menge{0,\ldots,N}$.
	\item Let $\Theta^+=\menge{\bbeta\in\Theta:p(\bbeta)>0}$. If $\Theta^+$ has finite Lebesgue measure and $p(\bbeta)$ and $q(\bbeta,\by)$ are bounded away from zero on $\Theta^+$, then there exists a constant $M$ and $r<1$ such that

		\[ \sup \menge{\Big| \bP^m(A|\bbeta^0)-\int_A{p(\bbeta)} \d \bbeta \Big|, {A\subset\Theta \text{ measurable}},\bbeta^0\in\Theta }  \leq M r^m,\]
		where $ \bP^m(\cdot|\bbeta^0)$ is the probability measure of the $m$th output of the Metropolis-Hastings algorithm with given $\bbeta^0$.
		
		Furthermore, for every function $f$ with $$\int_{\Theta} (1+\abs{f(\bbeta)})^2 \ p(\bbeta)\d\bbeta<\infty$$ there exists a constant $\sigma_f$, such that for $N\to\infty$
		\[ \sqrt{N}\left(\frac{1}{N}\sum_{m=1}^N f(\bbeta^m) - \int_\Theta  f(\bbeta) p(\bbeta) \d \bbeta \right) \xrightarrow {\;d\,} \nN(0,\sigma_f^2),  \]
		where $\nN(0,\sigma_f^2)$ is the normal distribution with variance $\sigma_f^2$ and the arrow indicates convergence in distribution.
		
\end{itemize}

Concerning the second issue, one should point out that the efficiency strongly depends on the proposal distribution $q(\cdot,\by)$. For instance if $q(\cdot,\by)\approx p$ the MHA will work most efficiently. Of course, usually this is not the case. However, it is noted in \cite{AuS01} that the efficiency does not depend on the type of the proposal distribution, but much more on the spread, which usually can be controlled by the variance of the proposal distribution. If the variance is too small, the sample gets too correlated. But if the variance is too large, the acceptance rate is too low and one has only very few accepted samples. So the question may as well be, ``what is the optimal acceptance rate for the MHA''.

One of the first papers which tries to give an answer to this question, is \cite{GelmA96}. The authors show that for a standard normally distributed target distribution the optimal acceptance rate is approximately $44 \%$ for dimension $d=1$ and $23 \%$ for dimension $d\rightarrow\infty$. The concept has been generalized to a multidimensional random variable with independent components in \cite{BedaM08}. In this case the number $23 \%$ for very high dimensions is not always attained, depending on the target distribution. The authors state that it remains still unclear what is an optimal acceptance rate for a general target distribution.

There is an additional issue concerning the efficiency of the MHA: the \textit{burn-in phase}. Like stated above, if the initial distribution is the target distribution, we have perfect sampling. However, if the initial distribution is far away from the target distribution, the first few output samples of the MHA are not distributed according to the target distribution, and thus one deletes the first few outputs of the algorithm. For the interested reader we refer to \cite{Brooks1998}, where several methods of output analysis (such as \emph{variance ratio method, spectral method or cumulative sum method}) are described. These methods can help to determine if the chain has converged to the stationary distribution.

In the case of the 1D telegraph equation, we are interested in the posterior distribution of the parameters $\btheta = (b,c)$. Since we do not have any prior knowledge about the parameters except reasonable bounds on their range, we assume the uniform distribution with bounds
\begin{align*}
b_\text{min}&=0.05  &b_\text{max}&=0.6\\
c_\text{min}&=0.2 &c_\text{max}&=0.25,
\end{align*}
which were chosen based on numerical experiments.

The choice of the features $\balpha$ is non-trivial. Unlike in the subsection before, one cannot take the square difference between the measured signal and some ``mean'' signal, since there is no such thing as a ``mean'' signal. If one takes the arithmetic mean of all measured signals, by the phase shift, this is likely to be close to zero. Thus, we chose a set of characteristic features of the signal. It turned out that the phase angle and the amplitude of the three most dominant (in terms of amplitude) frequencies of the first echo in the discrete Fourier spectrum  are feasible as model features.
These were computed as follows: We set the signal before and after the first echo to zero (i.e., between 11.8 $\mu$s and 22 $\mu$s) and compute the discrete Fourier transform. The frequencies are between $1.12$ and $1.19$ MHz (i.e. the $33$rd to $35$th entry).
So, the features were $\balpha_{\text{}}=(\phi_{33},\phi_{34},\phi_{35},r_{33},r_{34},r_{35})$ which are the amplitudes and the phase angles. We assume to have a normally distributed error with zero mean and covariance matrix $\Sigma$. In \cite[Chapter 2]{Tarantola2005} it is suggested to choose $\Sigma$ from measured data. Since we had an undamaged plate at hand, we used the scans of that plate to compute $\Sigma$. In fact, the features $(\widetilde{\balpha}_{1,\text{meas}},\ldots,\widetilde{\balpha}_{n,\text{meas}})$
obtained at the $n$ grid points of the undamaged plate can be seen as realizations of the feature values at any single location (assuming homogeneity of the underlying random field) and hence can serve to estimate the covariance matrix $\Sigma$.

Although the support of the probability density of the normal distribution is the whole space (contrary to the features) the likelihood function is chosen as
\[ f_{\bx|\btheta=\bbeta}(\balpha) = C \exp\left(-\frac{(\nM(\bbeta)-\balpha)^T {\Sigma^{-1}} (\nM(\bbeta)-\balpha)}{2} \right), \]
where $C$ is a constant and the function $\nM$ is the numerical solution operator from the previous section, computing the features $(\phi_{33},\phi_{34},\phi_{35},r_{33},r_{34},r_{35})$ for given parameters $b$ and $c$.  The choice can be justified since the variances were very small and thus $f_{\bx|\btheta=\bbeta}$ decreased fast enough and no further cut-offs were necessary.

So for the $i$th measurement location we have
\[ %
\begin{aligned}
&f_{\btheta|\bx=\balpha_{i,\text{meas}}}(\bbeta) \\
=&D \exp\left(-\frac{(\nM(\bbeta)-\balpha_{i,\text{meas}})^T {\Sigma^{-1}} (\nM(\bbeta)-\balpha_{i,\text{meas}})}{2} \right) \1_{[0.1,0.6]\times[0.2,0.25]}(\bbeta),
\end{aligned}
\]
where $D$ is a constant.

\begin{figure}[tbh]
	\centering
	\raisebox{5mm}{(a)}\quad \includegraphics[width=0.355\linewidth]{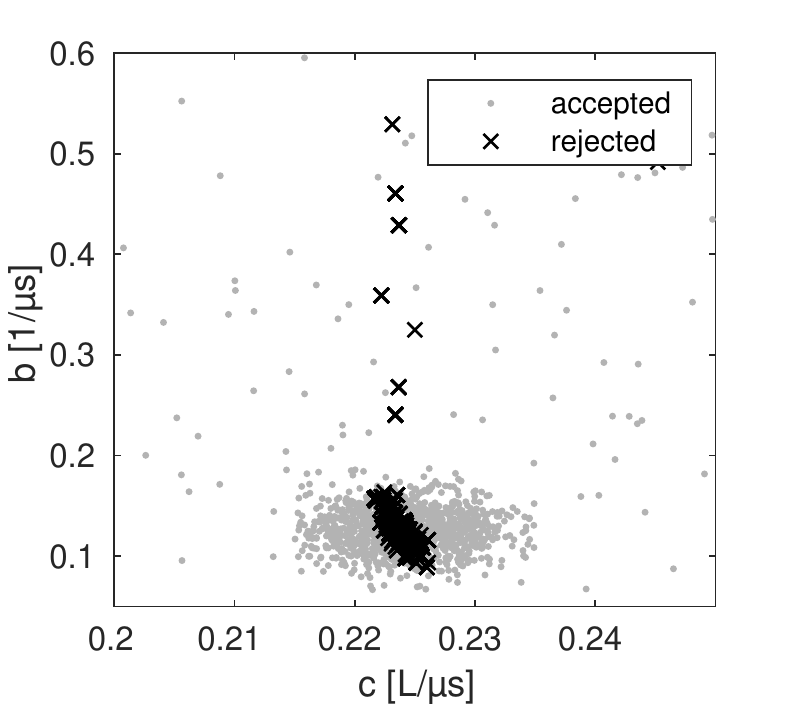} \quad
    \raisebox{5mm}{(b)}\quad \includegraphics[width=0.44\linewidth]{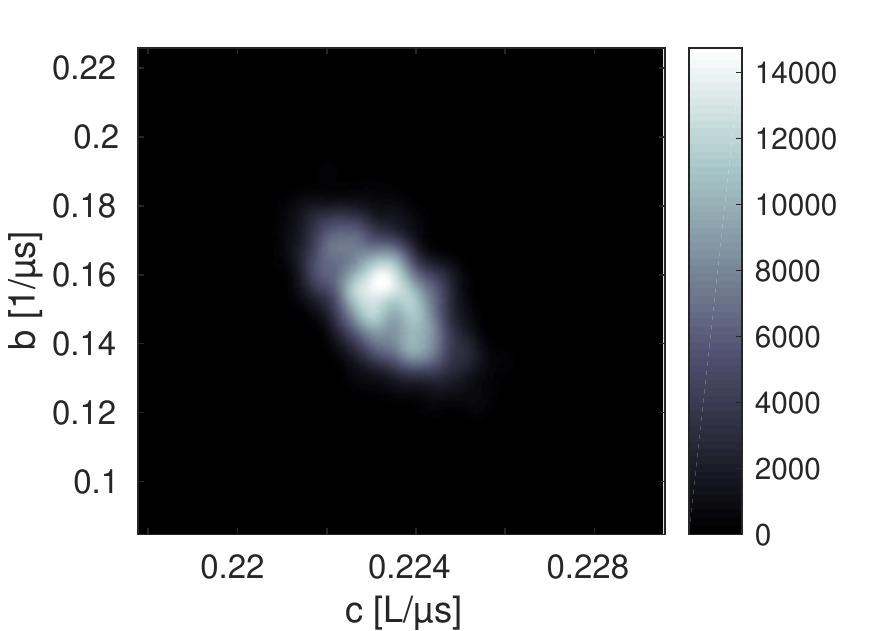}
	\caption{(a) MHA generated sample with initial guess $c_0=0.225$ and $b_0=0.35$. After the burn-in phase, the chain settles in the region around $c\approx0.224$ and $b\approx0.12$; (b) smooth kernel density estimate of the joint posterior distribution. The gray scale in (b) indicates the value of the probability density}
	\label{fig:mhavis}
\end{figure}

The Metropolis-Hastings algorithm was performed with the following proposal distribution:
\[ \boldeta^k \sim \nN(\bbeta^{k-1},\Sigma^{\text pr}_{k-1}),\]
where
\[ \Sigma^{\text pr}_k=\epsilon_k
\begin{bmatrix}
(b_{\text{min}}-b_{\text{max}})&0\\
0& (c_{\text{min}}-c_{\text{max}})
\end{bmatrix}\\
\]
and
\[ \epsilon_k=\begin{cases}
0.02 & \text{if } k< 100\\
0.001& \text{else}.
\end{cases} \]
Experiments with the model showed that the posterior distribution of $b$ and $c$ is concentrated in the  center of the prior domain. So we set the initial guess at $\bbeta^0=\frac{1}{2}({b_{\text{min}}+b_{\text{max}}},{c_{\text{min}}+c_{\text{max}}})$. For this setup the condition of an exponential convergence of the Markov chain to the target distribution is satisfied. Since $\bbeta^0$ is not $f_{\btheta|\bx}$-distributed, we neglect the first $100$ samples of the algorithm as we interpret this as burn-in phase. After the burn-in phase, the length of the Markov chain was $N=1000$. In Figure \ref{fig:mhavis}(a) one can see a typical example of the Markov chain at one certain measurement location. After the burn-in phase it settles in a certain region and stays there. Figure \ref{fig:mhavis}(b) shows the joint posterior distribution of $b$ and $c$ at this measurement location.

\begin{figure}[tbh]
	\centering
	\raisebox{6mm}{(a)} \includegraphics[width=0.4\linewidth]{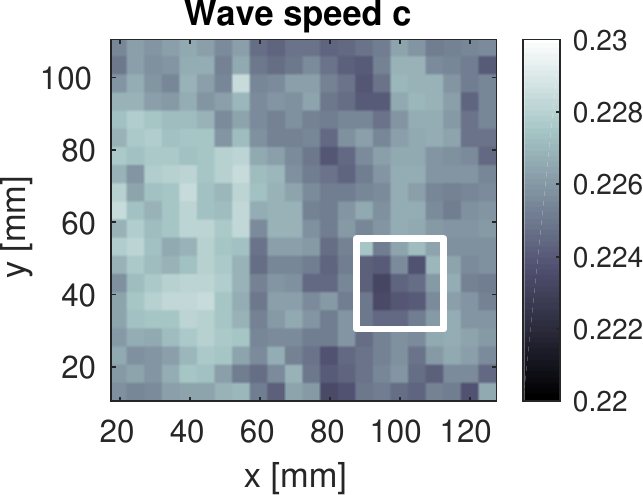} \qquad
    \raisebox{6mm}{(b)} \includegraphics[width=0.38\linewidth]{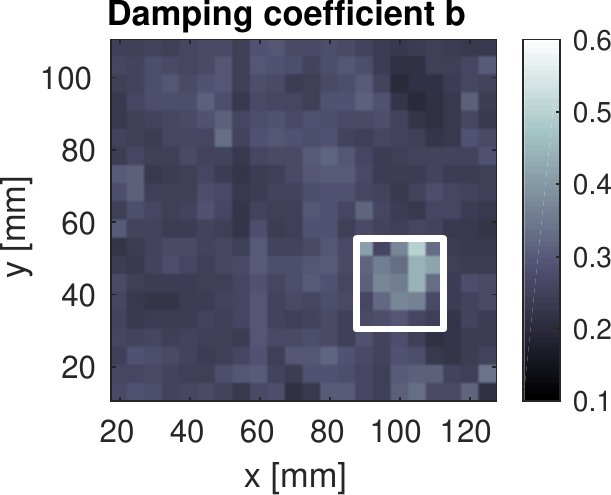}\\[6pt]
    \hspace{7pt} \raisebox{6mm}{(c)} \includegraphics[width=0.4\linewidth]{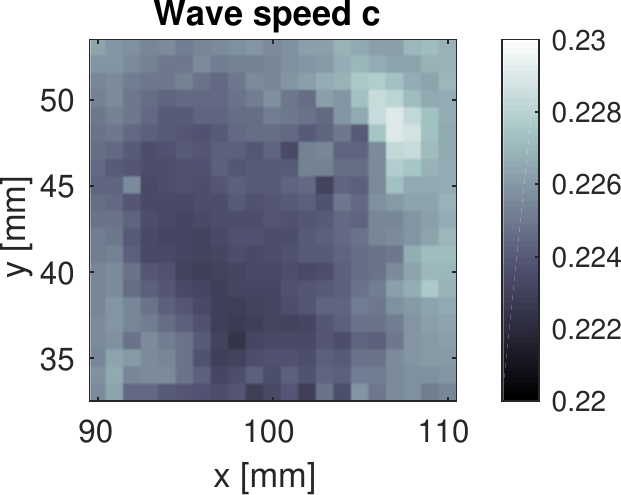} \qquad
    \raisebox{6mm}{(d)} \includegraphics[width=0.38\linewidth]{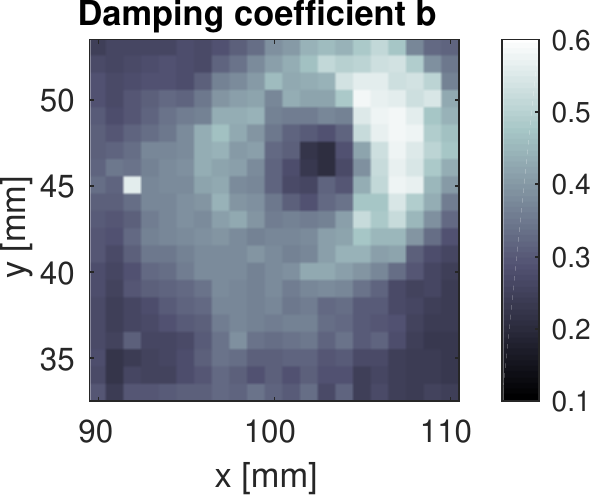}
	\caption{Spatial resolution of the posterior distribution's mean value: (a) overall view of wave speed $c$; (b) overall view of damping parameter $b$; (c) and (d) high resolution of damaged region. The gray scale is in units L/$\mu$s for $c$ and in units 1/$\mu$s for $b$. The $x$- and $y$-coordinates refer to the location on the positioning platform.}
\label{fig:platte3_mean}
\end{figure}

Figure \ref{fig:platte3_mean} depicts the spatial resolution of the mean value of the posterior distribution in each scan location. One can observe that the image is not as clear as in Figure \ref{fig:platte3}. This comes from the fact that the chosen features were only the phase angle and amplitude of the most dominant frequencies of the first echo. One may get different results with other features.

\subsection{Bayesian damage test}
\label{subsec:BayesTest}

We now formulate the following Bayesian hypothesis test (see e.g. \cite[Chapter 8, §1]{Gelman2004}). As null hypothesis we assume that the material is undamaged. This means that the material parameters $(b,c)$ are in the domain
\[\Theta_0:=\menge{(b,c)\in\Theta, b<b_{\text{crit}}, c>c_{\text{crit}}},\]
defined by critical thresholds $b_{\text{crit}}$, $c_{\text{crit}}$.
Then the posterior probability of the null hypothesis is just
\[ \PP(\btheta\in\Theta_0|\bx=\balpha) =\int_{\Theta_0} f_{\btheta|\bx=\balpha}(\bbeta) \d \bbeta. \]
We chose a confidence level of $1 \%$ and, thus, the null hypothesis is rejected if $\PP(\btheta\in\Theta_0|\bx=\balpha)<0.01$.

In real life applications, the thresholds $b_{\text{crit}}$, $c_{\text{crit}}$ will be given by engineering requirements on the material properties of the plates. Alternatively, they could be estimated from the response of an undamaged plate.
In our case, the threshold $b_{\text{crit}}$ resp. $c_{\text{crit}}$ was chosen from the optimized parameters (cf. Subsection \ref{subsec:optim}) such that $99 \%$ parameters of the undamaged region of the same plate were below $b_{\text{crit}}$ resp. above $c_{\text{crit}}$. The pragmatic reason for this choice was that we could not use the estimated parameters from another, undamaged plate, since all plates were subjected to a grinding procedure which resulted in a thickness difference of $2-3 \%$ between different plates. Due to our modelling assumption of constant plate thickness $L$, taking data from different plates would produce an error of the wave speed in the same range, and thus would affect the choice of $c_{\text{crit}}$. For that reason we used the undamaged region of the damaged plate as reference.

\begin{figure}[tbh]
	\centering
	\raisebox{6mm}{(a)}\quad \includegraphics[width=0.42\linewidth]{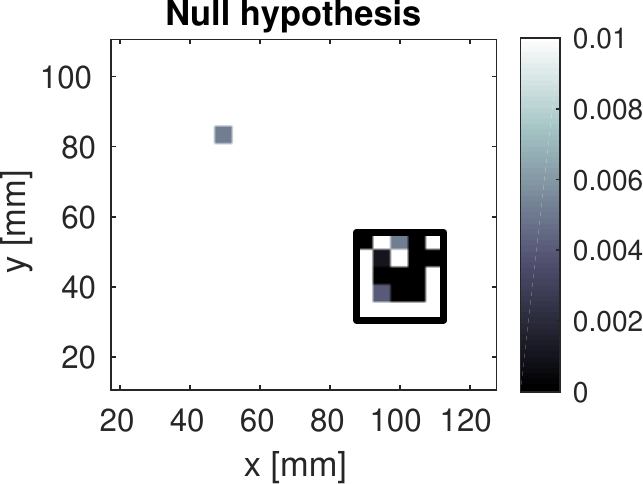} \quad
    \raisebox{6mm}{(b)}\quad \includegraphics[width=0.39\linewidth]{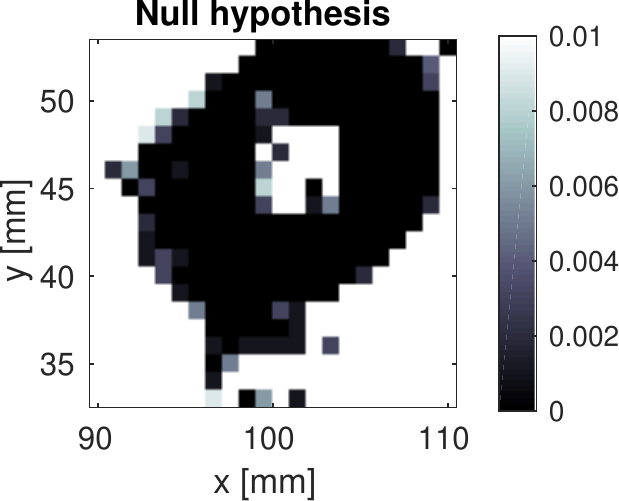}
	\caption{Spatial resolution of the posterior probability of the null hypothesis. In the white region the null hypothesis is not rejected. The gray scale indicates probability; (a) overall view; (b) high resolution of the damaged region. The $x$- and $y$-coordinates refer to the location on the positioning platform.}
	\label{fig:platte3pvalue}
\end{figure}

Figure \ref{fig:platte3pvalue} shows the spatial resolution of the posterior probability of the null hypothesis. It is seen that the null hypothesis is rejected in the area enclosed by the black square, which coincides with the actually damaged area of the plate under investigation (compare with Figure \ref{fig:experiment}), and the null hypothesis is accepted in the undamaged parts. Thus the damage is correctly identified, at least within the accuracy admitted by the grid size.

\section{Discussion}

As outlined in the introduction, a wealth of methods for damage detection based on ultrasonic measurements have been developed and are in use. The approach presented in this paper is a refined analysis of data acquired by so-called A-scans \cite{Lammering2018,Krautkraemer1990}. An A-scan is a unidirectional measurement, usually in the direction perpendicular to the surface of the material. A change in the transit time of the reflected wave indicates a damaged area. It is common industrial practice to check the production quality, e.g., of carbon fiber composite plates, by rapidly performed A-scans. The intention of the present paper is to extract additional information by analyzing the time dependent, dynamic signal as recorded by the oscilloscope. This is done by setting up a simple model for one-dimensional wave propagation and calibrating wave speed and damping coefficient to the recorded data. This way two independent indicators are obtained, as compared to the usual practice of monitoring the transit time only.
An additional benefit of our approach is that our model can be solved analytically, thereby enabling a rapid computation of the exact solution. The low computational cost is a big advantage in the required optimization step or the evaluation of the posterior distribution, respectively. Using exact analytical solutions, when possible, has also been recommended in the literature \cite{Lammering2018}, \cite[Section 4]{Ostachowicz2010}.

Our analysis is solely based on the data recorded by the oscilloscope. It does not require knowledge of the absolute value of the plate thickness nor the value of the wave speed, so no further (mechanical or geometrical) measurements are needed (in the case of plates of constant thickness). On the other hand, it does not allow one to infer these quantities either. Another limitation is that our method does not admit the determination of the elastic parameters (the Lam\'e constants) of the material.

A possible application could be to automated scanning, where the decision about the presence of damage could be based on the hypothesis test presented in Subsection\;\ref{subsec:BayesTest}. Finally, there is also some theoretical interest in our approach: the mathematical model is extremely simple, yet it proved to be capable of describing A-scans and processing information needed for damage detection.

\section{Summary}

The paper addressed the possibility of parameter calibration and damage detection based on the time dependent response of a structure under acoustic excitation. The work extended from actual experimental impulse echo measurements to
establishing a mathematical model of wave propagation and reflection, solving the model equations by an efficient numerical procedure based on methods from Fourier analysis, and finally calibrating the model parameters by comparing the computed
response to (features) of the measured data. The calibration was done by deterministic optimization and by a Bayesian approach using the Metropolis-Hastings algorithm. The posterior distribution of the parameters can be used to design a hypothesis test detecting damage and its location on the structure. An analogous method for three-dimensional elastic solids has been worked out in \cite{Schwarz:Thesis:2019}, announced in \cite{MOMS:14} and will be the topic of a future publication.

\section*{Acknowledgements}

Foremost we wish to thank our industrial partner INTALES GmbH Engineering Solutions, Natters, Austria, for continuous support and advice and for providing the plate specimen. We thank Jonathan Halmen for carrying out the ultrasonic measurements and also for installing and programming the control of the positioning table. We thank Peter Paulini for providing the equipment and for valuable advice on ultrasonic flaw detection. The project was supported by grant No. 4602529, Bridge Program, The Austrian Research Promotion Agency (FFG), together with INTALES GmbH Engineering Solutions. The second author acknowledges support through grant P-27570-N26 of The Austrian Science Fund (FWF). The computational results presented have been achieved (in part) using the HPC infrastructure LEO of the University of Innsbruck.

\end{document}